\documentclass[cits]{PoS}

\title{Progress toward the chiral regime in lattice QCD calculations of
the neutron electric polarizability}

\ShortTitle{Progress toward the chiral regime in the neutron electric
polarizability}

\author{\speaker{Michael Engelhardt}%
         \thanks{Supported by the U.S.~DOE under grant DE-FG02-96ER40965.}\\
        Department of Physics, New Mexico State University, Las Cruces\\
        E-mail: \email{engel@nmsu.edu}}

\abstract{The static electric polarizability of the neutron is evaluated
using domain-wall valence quarks on a MILC asqtad sea quark ensemble
corresponding to a pion mass of $m_{\pi } =357\, \mbox{MeV} $. Both
connected as well as disconnected contributions are included. The result
is quantitatively compatible with the expectation from chiral effective
theory.}

\FullConference{The XXVII International Symposium on Lattice Field Theory\\
                 July 26-31, 2009\\
                 Peking University, Beijing, China}

\begin{document}

\section{Introduction}
Electromagnetic polarizabilities are fundamental properties of hadrons
which figure prominently, e.g., in soft Compton scattering. In the low
energy limit, when the wavelength of the scattered photons is large
compared to the hadron size, the Compton scattering amplitude can be
cast in terms of a gradient expansion in the electromagnetic field
\cite{polariz}. The leading terms in this expansion are determined by
the static electric and magnetic polarizabilities, which are defined
via the hadron mass shift in constant external electric and magnetic
fields, respectively. In this work, specifically the static electric
polarizability of the neutron is investigated, building on earlier
exploratory work \cite{elpol}. By pushing the calculation down to
a pion mass of $357\, \mbox{MeV} $, contact is established with chiral
effective theory; the approach is fully dynamical, including all
disconnected quark loop contributions. Related efforts have recently
been reported by several groups, cf.~\cite{alexandru,detmold,guerrero}.

\section{Expansion of the neutron two-point function}
\begin{figure}
\includegraphics[width=15cm]{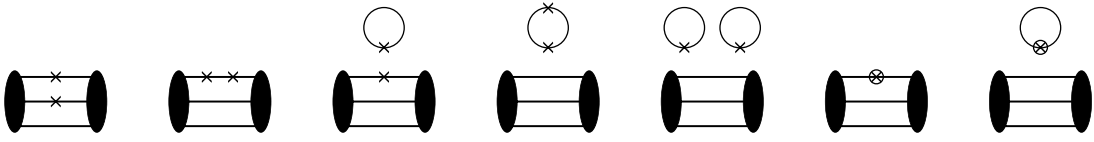}
\caption{Diagrammatic representation of contributions to the neutron
electric polarizability. Crosses denote interaction vertices linear in
the external field, circled crosses interaction vertices quadratic in
the same.}
\label{diagrams}
\end{figure}

In parallel to the Compton scattering amplitude alluded to above, the
response of the neutron mass to an external electric field
can be expanded both with respect to the magnitude of the electric field
as well as its space-time variation. Here, only the leading term in this
combined expansion is of interest,
\begin{equation}
m(E) -m(0) = -\frac{\alpha }{2} E^2 + \ldots \ ,
\label{mshift}
\end{equation}
discarding all space-time derivatives of the electric field and terms
higher than quadratic in the same\footnote{From the outset, the absence
of a neutron electric dipole moment is assumed in the present treatment;
this would of course no longer be accurate in the presence of a non-trivial
vacuum $\Theta $-angle.}. The electric polarizability $\alpha $ can then
be extracted by evaluating the neutron two-point function, which
determines the neutron mass, in the presence of a {\em constant} electric
field; the two-point function can furthermore be expanded in that field
from the outset, keeping only quadratic terms. Consider decomposing the
action into the zero external field term and the perturbation due to
the external field, $S=S_0 +S_{ext } $. Then the two-point function
expands as
\begin{eqnarray}
\left\langle N_{\beta } (y) \bar{N}_{\alpha } (x) \right\rangle &=&
\left. \int [DU] [D\psi ] [D\bar{\psi } ] \ \exp (-S_0 ) \
\left( 1 - S_{ext} + S_{ext}^{2} /2 + \ldots \right)
N_{\beta } (y) \bar{N}_{\alpha } (x) \ \right/ \nonumber \\
& & \hspace{2cm} \int [DU] [D\psi ] [D\bar{\psi } ] \ \exp
(-S_0 ) \ \left( 1 - S_{ext} + S_{ext}^{2} /2 + \ldots \right)
\label{twopt}
\end{eqnarray}
As usual, the terms in the denominator cancel statistically disconnected
parts of diagrammatically disconnected contributions. The coupling to
the external electric field still remains to be specified. In the following,
the electric field will be taken to point into the spatial 3-direction,
and will be introduced via the 3-component of the gauge field,
$A_3 =E(t-t_0 )$. This choice will be discussed in depth further
below; it decisively influences the physics of the external field
method calculation pursued here. For now, it is sufficient to remark
that only $A_3 $ is non-zero, that it is constant in the 3-direction,
and that it is linear in $E$. Under these circumstances, the gauge
links are modified by the external field as
\begin{equation}
U_3 \longrightarrow
\exp \left( iq \int dx_3 \cdot A_3 \right) \cdot U_3
= \left( 1 + iaqA_3 - a^2 q^2 A_3^2 /2 + \ldots \right) \cdot U_3
\end{equation}
(where $a$ denotes the lattice spacing and $q$ the quark electric
charge matrix)
and, inserting this into fermion action discretizations of the Wilson
type, the coupling to the external field takes the form\footnote{The
renormalization factor $Z_V $ is due to a technical detail: In the
numerical calculation, quark lines in Fig.~\ref{diagrams} will be
populated by domain wall fields; however, the coupling to the
external field is not realized via the five-dimensional conserved
current, but via the current obtained by projecting the quark modes
onto the four-dimensional domain walls. The fields $\bar{\psi } $, $\psi $
in (\ref{vertices}) are the projected ones. This amounts to a
renormalization of the vertices, which is compensated by $Z_V $. The
renormalization factor is determined to be $Z_V =1.14$ for light quarks
and $Z_V =1.10$ for strange quarks by measuring the number of valence
quarks in baryons.}
\begin{eqnarray}
S_{ext} &=& Z_V \cdot \frac{1}{2} \sum_{x} \ \bar{\psi } (x) \left[
(iaqA_3 - a^2 q^2 A_3^2 /2) \cdot U_3 (x) \cdot (-1+\gamma_{3} ) 
\cdot \psi (x+e_3 ) ) \right. \label{vertices} \\
& & \ \ \ \ \ \ \ \ \ \ \ \ \ \ \ \ \ \ \ \ \ \ \ \ \ \ \ \ \ \left.
+ \ (iaqA_3 + a^2 q^2 A_3^2 /2) \cdot U^{\dagger }_{3} (x-e_3 ) \cdot
(1+\gamma_{3} ) \cdot \psi (x-e_3 ) ) \right]
\nonumber
\end{eqnarray}
Thus, two interaction vertices are generated, one linear in the
external field, and one quadratic. Carrying out the fermionic
integrations in (\ref{twopt}), i.e., applying Wick's theorem,
leads to these vertices being inserted into quark propagators,
yielding the diagrammatic representation of the relevant contributions
to the neutron two-point function depicted in Fig.~\ref{diagrams}.

\section{Role of the external gauge field}
\label{extsec}
As already mentioned in the previous section, the external electric
field will be introduced via a non-vanishing 3-component of the gauge
field,
\begin{equation}
A_3 =E(t-t_0 ) \equiv A+Et \ .
\label{a3def}
\end{equation}
To arrive at a cogent interpretation of the neutron two-point function
measured in the presence of such a field, a detailed consideration
of the resultant physical effects is called for. First and foremost,
note the freedom in the choice of the origin in the time direction,
$t_0 $, which is equivalent to a freedom of adding a constant, $A$,
to the gauge field. If the spatial directions were infinite, this
choice would be inconsequential, since $A$ could be freely shifted
by a gauge transformation. However, on a finite space of linear extent
$L$ with periodic boundary conditions, only a discrete gauge invariance
$A\longrightarrow A+2\pi /q_0 L$ remains, where $q_0 =e/3$ represents the
elementary electric charge present in the system. Apart from this residual
invariance, physics, and, in particular, the mass shift of the neutron
depends on the chosen value of $A$. The question therefore arises how the
electric polarizability proper can be disentangled from the $A$-dependence
of the neutron mass.

Physically, $qA$ enters the quark Hamiltonian\footnote{In more precise
terms, this would be an effective constituent quark Hamiltonian;
the electrical current itself depends on the effective mass experienced
by the charge carriers as they propagate with a given Bloch momentum.}
as a quark Bloch momentum in the periodic lattice. Thus, for general $A$,
there is a current of up quarks flowing around the torus with each up
quark, on the average, carrying momentum $2eA/3$, and a current of twice
as many down quarks with, on the average, half as much momentum each in the
opposite direction\footnote{In addition to these valence currents,
there are, of course, also analogous sea quark and antiquark currents.}.
Only the complete (periodic) ``neutron'', if it can be viewed as such,
maintains vanishing momentum. Of course, the notion of a neutron becomes
tenuous when the Bloch momenta become large; in this limit, the system
essentially consists of two strong, oppositely oriented quark electrical
currents with some residual correlations due to the chromodynamic force.
Certainly, any mass shift extracted in such a regime could hardly be
construed as representing the polarizability of a reasonably isolated
neutron. Instead, it is necessary to seek out the regime where the quark
currents cease and a bona fide neutron, only weakly distorted by the
electric field $E$ proper, is formed.

In the absence of an actual electric field, it is clear how to achieve
this situation; one must set $A=0$. When $E\neq 0$, on the other hand,
the situation is more complicated, since the Bloch momenta change (linearly)
in time: In view of (\ref{a3def}), a shift in the time direction is
equivalent to a shift in $A$. This is, of course, entirely natural;
the electric field accelerates the up and down quark electrical currents.
A practical and model-independent way to identify when the quark currents
cease is to look for stationarity in physical quantities such as the
mass shift. If the quark currents are initially oriented opposite to
the acceleration provided by the electric field, then time evolution
will lead to a point at which the currents cease and turn around. 
Expanded around that point in time, the leading time dependence of
physical quantities is quadratic. Equivalently, one may perform
measurements of the neutron two-point function in a fixed time
region, but for different values of $A$, and subsequently seek out
the stationary point as a function of $A$. In practice, this calls
for measurements at three different $A$, permitting the determination
of a parabola in that variable, as well as its stationary point.

Before continuing the main line of discussion by turning to the
extraction of the neutron mass shift from its two-point function,
cf.~section~\ref{section4}, a separate concern associated with the choice
(\ref{a3def}) of the external gauge field should be addressed.
While this gauge field is constant, and thus periodic in the spatial
directions, it must exhibit a discontinuity at some point in the
time coordinate in order to be periodic in that direction as well.
At the discontinuity, the electric field exhibits a spike, with the
potential of distorting the measurement of the neutron mass shift, which
is, after all, intended to be taken in a constant electric field. As
demonstrated in \cite{detmold}, such an effect can indeed be seen
if one insists on performing the mass measurement in a time region
including the spike. For selected magnitudes of the electric field,
the distortion can be avoided, because what is physically relevant is
not the gauge field $A_3 $ itself, but $e^{iqA_3 } $; thus, if one
arranges the discontinuity to be an appropriate multiple of $2\pi $,
it has no physical consequences. However, ultimately,
this issue is moot, since nothing forces one to perform measurements
near the gauge field discontinuity\footnote{Note, however, that for the
alternative choice of gauge field $A_0 = Ex_3 $, the analogous objection
would indeed be more serious; during the mass measurement, one cannot
prevent the neutron from propagating to the spatial location of the
electric field spike. Indeed, if one projects onto definite neutron
momentum, all spatial positions are explicitly weighted equally.}.
In fact, in practice, it is not even necessary to extend the support of
$A_3 $ over all times; in the present investigation, just as in a real
laboratory setting, the external field is merely switched on a little
before the neutron source is introduced, and switched off some time after
it is annihilated. This does not alter the result of the mass measurement,
as tested in \cite{elpol}; however, it significantly reduces the statistical
fluctuations in disconnected diagrams, since quark loops far in the past
and the future of (and thus not correlated with) the neutron are eliminated.

\section{Extracting the mass shift from the two-point function}
\label{section4}
As should be clear from the discussion in the previous section, the neutron
Hamiltonian in the presence of the external field adopted here is not
time-independent; a shift in time corresponds directly to a shift in the
constant $A$ in the external field (\ref{a3def}). Before addressing the
time dependence, it is useful to review how the mass shift would enter
the neutron two-point function in the case of a time-independent
Hamiltonian depending on two arbitrary small parameters $E$ and $A$.
In such a case, projecting onto zero-momentum, unpolarized neutrons,
the two-point function behaves as
\begin{equation}
G(p=0,t) =
\sum_{\vec{y} } \ \mbox{Tr} \left( \frac{1+\gamma_{0} }{2}
\left\langle N (y) \bar{N} (x) \right\rangle \right)
\ \ \stackrel{t\rightarrow \infty }{\longrightarrow } \ \
W \ \exp (-m \, t)
\label{decay}
\end{equation}
for large times, where both the overlap between neutron source and true
ground state $W$ and the neutron mass $m$ can be expanded in $E$ and $A$,
\begin{eqnarray}
W &=& W_0 + W^{(1)} (A,E) + W^{(2)} (A,E) +\ldots \\
m &=& m_0 + m^{(2)} (A,E) +\ldots
\end{eqnarray}
Inserting this into (\ref{decay}), the term of second order in $E$ and $A$
in the two-point function becomes
\begin{equation}
G^{(2)} (p=0,t) \ \ \stackrel{t\rightarrow \infty }{\longrightarrow } \ \
W_0 \ \exp (-m_0 \, t) \left( \frac{W^{(2)} (A,E)}{W_0 }
-m^{(2)} (A,E) t \right)
\label{order2}
\end{equation}
implying that the neutron mass shift to second order in $E$ and $A$ can be
obtained from (minus) the temporal slope of the correlator ratio
$R(t)=G^{(2)} (t)/G^{(0)} (t)$.
\vspace{0.3cm}

\begin{figure}[h]
\hspace{1cm}
\includegraphics[angle=-90,width=5cm]{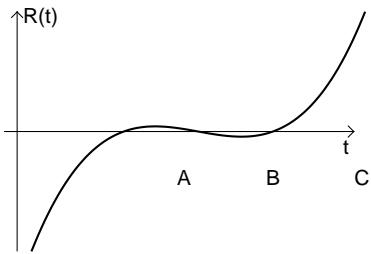}
\vspace{-0.01cm}
\caption{\hspace{0.1cm} Generic \hspace{0.1cm} behavior \hspace{0.1cm} of
\hspace{0.1cm} the \hspace{0.1cm} correlator \hspace{0.1cm} ratio
\hspace{6.74cm} $\mbox{\ \ } $
$R(t)=G^{(2)} (t)/G^{(0)} (t)$. \hspace{0.1cm} Regions \hspace{0.1cm} 
A, \hspace{0.03cm} B \hspace{0.03cm} and \hspace{0.03cm} C 
\hspace{0.03cm} qualitat- \hspace{6.75cm} $\mbox{\ \ } $ ively
correspond to the time windows displayed in Fig.~3.}
\label{cubic}
\end{figure}
\vspace{-6.45cm}

\hspace{8cm} \parbox{6.32cm}{
Now, the Hamiltonian of interest here is not time-independent; however,
if it is sufficiently weakly time-dependent, the system will adjust
adiabatically while $W^{(2)} $ \hfill and \hfill $m^{(2)} $ \hfill
vary \hfill slowly, \hfill and \hfill (\ref{order2}) \hfill still}
\vspace{-0.09cm}

\hspace{7.98cm}
applies\footnote{A treatment beyond the adiabatic approximation,
in the related context of the acceleration of a proton in an external
electric field, has been described in \cite{detmold}.}, \hfill
except \hfill that \hfill $W^{(2)} $ \hfill and \hfill $m^{(2)} $
\hfill are \linebreak $\mbox{\ \ } $ \vspace{-0.3cm}

\hspace{8cm} \parbox{6.32cm}{
now time-dependent.  This implies that $m^{(2)} $ cannot
be simply read off as the slope of $R(t)$ anymore; this
correlator ratio generically will behave as displayed in Fig.~\ref{cubic}.
However, as discussed further above, to obtain the electric
polarizability of \hfill the \hfill neutron, \hfill one \hfill in \hfill
particular \hfill needs \hspace{-0.2cm}
\vspace{0.13cm}
}

\pagebreak
\noindent
only the mass shift
at the stationary point in $A$, or, equivalently, $t$. Expanded around
that point, the time dependence of $W^{(2)} $ and $m^{(2)} $ is
quadratic; therefore, at that point, and only there, the mass shift
$m^{(2)} $ is still given by (minus) the slope of $R(t)$, thus permitting
the extraction of the neutron electric polarizability.

\section{Results}
\begin{figure}[t]
\includegraphics[angle=-90,width=6cm]{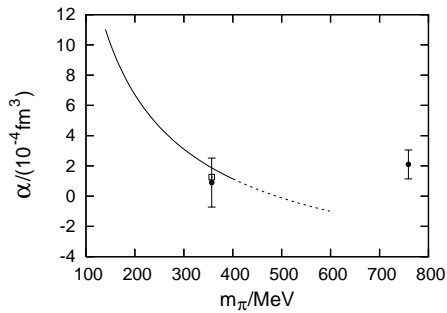}
\caption{Electric polarizability of the neutron;
\hspace{8.5cm} $\mbox{\ \ } $
solid \hspace{0.077cm} circles \hspace{0.077cm} with \hspace{0.077cm} error
\hspace{0.077cm} bars \hspace{0.077cm} denote \hspace{0.077cm} lattice
\hspace{8.5cm} $\mbox{\ \ } $
measurements \hspace{0.021cm} from \hspace{0.021cm} the \hspace{0.021cm}
present \hspace{0.021cm} work \hspace{0.021cm} (at \hspace{0.021cm}
$m_{\pi } $ \hspace{8.5cm} $\mbox{\ \ } $
$=357\, \mbox{MeV} $) \hspace{-0.023cm} and \hspace{-0.023cm} from
\hspace{-0.023cm} \cite{elpol} \hspace{-0.023cm} (at \hspace{-0.023cm}
$m_{\pi } =759\, \mbox{MeV} $). \hspace{8.5cm} $\mbox{\ \ } $
The \hspace{-0.021cm} curve \hspace{-0.021cm} represents \hspace{-0.021cm}
the \hspace{-0.021cm} SSE \hspace{-0.021cm} expectation \hspace{-0.021cm}
\cite{hgrie}, \hspace{8.5cm} $\mbox{\ \ } $
constructed \hspace{0.085cm} purely \hspace{0.085cm} from \hspace{0.085cm}
experimental \hspace{0.085cm} data. \hspace{8.5cm} $\mbox{\ \ } $
Apart \hspace{0.006cm} from \hspace{0.006cm} the \hspace{0.006cm} pion
\hspace{0.006cm} mass, \hspace{0.006cm} which \hspace{0.006cm} is
\hspace{0.006cm} varied \hspace{8.2cm} $\mbox{\ \ } $
in \hspace{-0.026cm} the \hspace{-0.026cm} graph, \hspace{-0.026cm} the
\hspace{-0.026cm} SSE \hspace{-0.026cm} expression \hspace{-0.026cm}
contains \hspace{-0.026cm} a \hspace{8.2cm} $\mbox{\ \ } $
collection \hspace{-0.009cm} of \hspace{-0.009cm} further \hspace{-0.009cm}
parameters, \hspace{-0.009cm} among \hspace{-0.009cm} them
\hspace{8.5cm} $\mbox{\ \ } $
the \hspace{-0.131cm} nucleon \hspace{-0.131cm} and \hspace{-0.131cm} delta
\hspace{-0.131cm} masses, \hspace{-0.131cm} $f_{\pi } $ \hspace{-0.131cm}
and \hspace{-0.131cm} $g_A $, \hspace{-0.131cm} cf.
\hspace{7.9cm} $\mbox{\, } $ 
\cite{hgrie}.~$\ $~For~$\ $~these \hspace{0.092cm}
parameters, \hspace{0.092cm} lattice \hspace{0.092cm} measure-
\hspace{8.5cm} $\mbox{\ \ } $
ments \hspace{-0.045cm} on \hspace{-0.045cm} the \hspace{-0.045cm} same
\hspace{-0.045cm} ensemble \hspace{-0.045cm} at \hspace{-0.045cm}
$m_{\pi } =357\, \mbox{MeV} $ \hspace{8.5cm} $\mbox{\ \ } $
exist; \hspace{0.023cm} substituting \hspace{0.023cm} those
\hspace{0.023cm} values \hspace{0.023cm} into \hspace{0.023cm} the
\hspace{0.023cm} SSE \hspace{8.5cm} $\mbox{\ \ } $
expression \hspace{0.013cm} yields \hspace{0.013cm} the \hspace{0.013cm}
open \hspace{0.013cm} square. \hspace{0.013cm} This \hspace{0.017cm} can
\hspace{8.15cm} $\mbox{\ \ \ } $
be \hspace{0.033cm} viewed \hspace{0.033cm} as \hspace{0.033cm} the
\hspace{0.033cm} result \hspace{0.033cm} of \hspace{0.033cm} an
\hspace{0.033cm} alternative, \hspace{8.18cm} $\mbox{\ \ } $
partially~resummed~effective~theory.
}
\label{eft}
\end{figure}

Numerical measurements were carried out in a mixed action scheme, using $448$
dynamical asqtad quark configurations provided by the MILC collaboration
\cite{milc}, with lattice spacing $a=0.124\, \mbox{fm} $ and quark masses
$am_l =0.01$, $am_s =0.05 $, corresponding to a pion mass of
$357\, \mbox{MeV} $. Quark lines depicted in Fig.~\ref{diagrams} were
populated with domain wall quarks; all disconnected contributions were
taken into account and were estimated using bulk complex $Z(2)$ stochastic
sources.
\begin{figure}[t]
\hspace{-0.7cm}
\includegraphics[angle=-90,width=5.7cm]{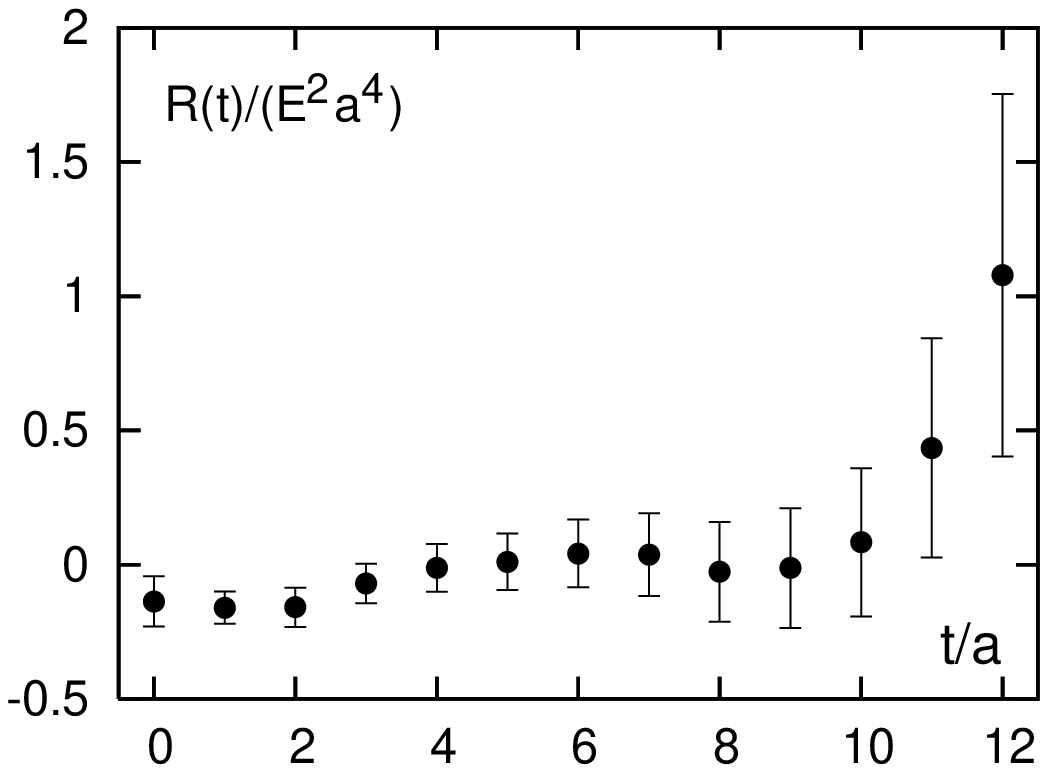} \hspace{-0.83cm}
\includegraphics[angle=-90,width=5.7cm]{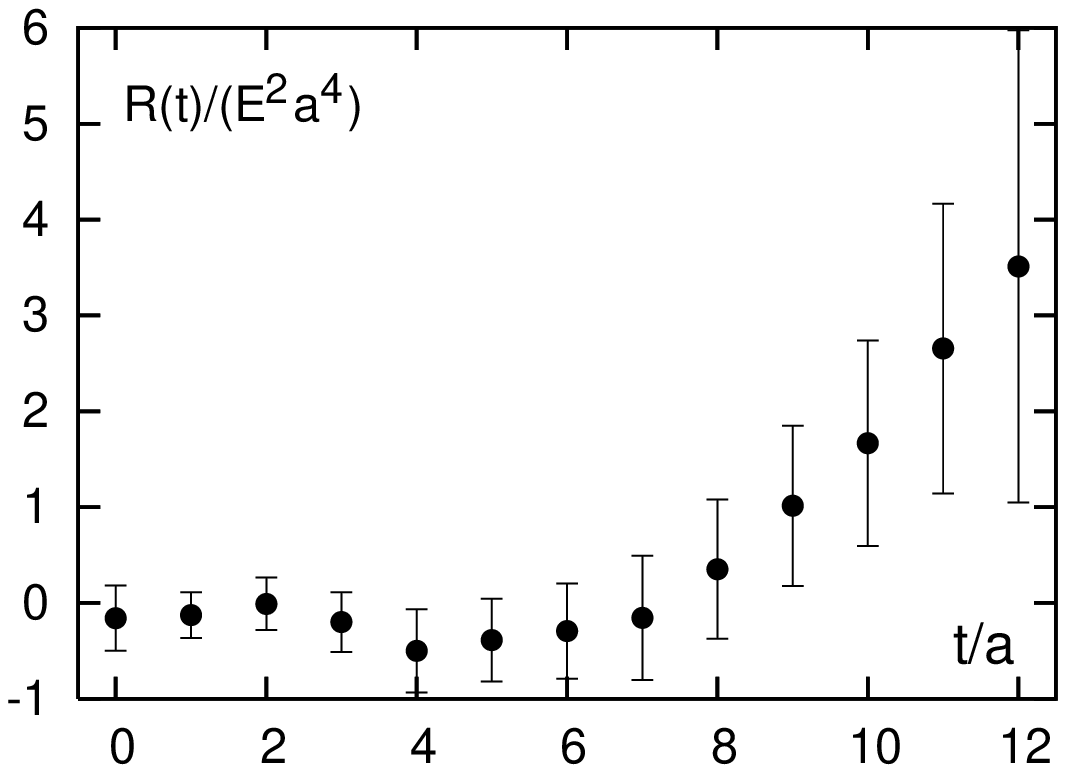} \hspace{-0.83cm}
\includegraphics[angle=-90,width=5.7cm]{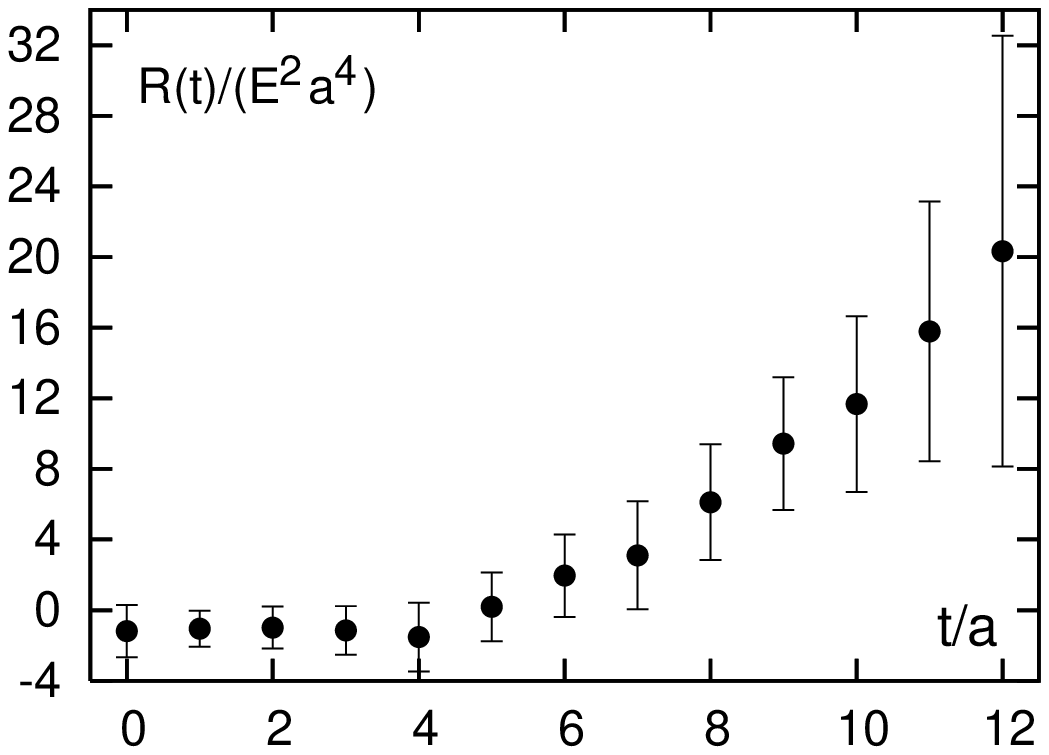} \hspace{-0.7cm}
\vspace{-0.3cm}
\caption{Correlator ratio $R(t)$ in (Gaussian) units of $E^2 a^4 $; in
these units, the plots are independent of any actual value of $E$ used
in the numerical computation. In the three measurements, the neutron
source was placed at the fixed location $t=0$, and the background
gauge field $A_3 =E(t-t_0 )$ was varied, with, from left to right,
$t_0 =6a$, $t_0 =0$ and $t_0 =-10a$. This generates time windows
corresponding qualitatively, from left to right, to regions A, B and C
in Fig.~2, respectively; note the variation in the vertical scale. The
behavior indicated in Fig.~2 is thus demonstrated, with the inflection
point, corresponding to the stationary point of the neutron Hamiltonian,
being captured in the left panel.}
\label{numer}
\end{figure}
Fig.~\ref{numer} displays three different measurements of the correlator
ratio $R(t)=G^{(2)} (t) /G^{(0)} (t) $ performed with different shifts $A$,
thus probing different time windows of the behavior displayed in
Fig.~\ref{cubic}.
As a result of using Wuppertal smeared neutron sources and sinks optimized
with respect to their overlap with the true neutron ground state,
as constructed in \cite{dolgov}, excited states are already
negligible for $t\ge 4a$ (where $t=0$ denotes the neutron source location).
On the other hand, statistical uncertainties are under reasonable control
up to $t=10a$. In practice, the average slope of the correlator ratio $R(t)$
for $5a\le t \le 9a$ was determined in each of the three time windows
displayed in Fig.~\ref{numer}; these three values define a parabola as a
function of $A$, the extremum of which yields (minus) the mass shift at the
stationary point, $-m^{(2)} (A_{stat} ,E)$. The latter was corrected for
curvature effects averaged over when determining the slope in each window.
Alternative analysis schemes were explored and yield consistent results.
Finally, supplementing $[-m^{(2)} (A_{stat} ,E)]/(E^2 a^3 )$ with a factor
$2$ from (\ref{mshift}) and a factor $-1$ from the continuation
$E\rightarrow iE$ to Minkowski time yields the polarizability, in
Gaussian units,
\begin{equation}
\alpha = (0.9\pm 1.6) \cdot 10^{-4} \mbox{fm}^{3} \ .
\label{result}
\end{equation}
Fig.~\ref{eft} puts this result in relation to chiral effective theory,
specifically the small scale expansion (SSE) \hfill expectation \hfill
\cite{hgrie}, \hfill with \hfill which \hfill it \hfill is \hfill 
quantitatively \hfill compatible. \hfill Although \hfill the \hfill
statistical \hfill uncer-

\hspace{6.5cm} \parbox{7.8cm}{\vspace{-21.9cm}
tainty
in (\ref{result}) is sizeable due mainly to the disconnected contributions,
the present calculation thus establishes contact with the pion mass
regime in which chiral effective theory becomes applicable;
an earlier
exploratory study \cite{elpol} at $m_{\pi } =759\, \mbox{MeV} $, the
\hfill result \hfill of \hfill which \hfill is \hfill also \hfill
included \hfill in \hfill Fig.~\ref{eft}, \hfill did}
\vspace{-10.18cm}

\hspace{6.5cm} not yet allow for a quantitative
comparison\footnote{As already explained in a note added in \cite{elpol},
the result for the polarizability initially quoted there was missing
an overall minus sign stemming from the continuation $E\rightarrow iE$
to Minkowski space.}.
\vspace{0.15cm}

\hspace{6.5cm} \parbox{7.8cm}{\hspace{0.5cm}
It is again worth emphasizing that, to
arrive at the result (\ref{result}), it was crucial
to carefully take into account the physics associated with an external
gauge field of the form (\ref{a3def}). As explained in detail in section
\ref{extsec}, on a finite spatial lattice, the external field induces
opposite Bloch momenta for the $u$- and $d$-quarks, with a bona fide
neutron only having an opportunity to form at a specific time, namely the
stationary point at which the Bloch currents reverse direction under the
influence of the electric field. That is where the electric polarizability
can be extracted. The measurements displayed in Fig.~\ref{numer} illustrate
how, in a background field calculation of this type, the polarizability
emerges as a rather small residual effect compared to the substantial
Bloch momentum dependence.
}

\acknowledgments
NMCAC is acknowledged for providing the computing resources for this work
on Encanto.

\end{document}